\RequirePackage{fix-cm}
\documentclass[smallextended]{svjour3}       
\smartqed  
\usepackage{graphicx}
\usepackage{color} 
\usepackage{mathtools}

\begin{document}

\title{Two-particle indistinguishability and identification of boson-and-fermion species: A Fisher information approach
}


\author{{Su-Yong~Lee} \and {Jeongho~Bang} \and {Jaewan~Kim}}


\institute{S.-Y.~Lee${}^\star$ (\email{papercrane79@gmail.com}) \and J.~Bang${}^\star$ (\email{jbang@kias.re.kr}) \and J.~Kim \at
              School of Computational Sciences, Korea Institute for Advanced Study, Hoegi-ro 85,Dongdaemun-gu, Seoul 02455, Korea \\
              ${}^\star$ Corresponding authors
}

\date{Received: date / Accepted: date}

\maketitle

\newcommand{\bra}[1]{\left<#1\right|}
\newcommand{\ket}[1]{\left|#1\right>}
\newcommand{\abs}[1]{\left|#1\right|}
\newcommand{\expt}[1]{\left<#1\right>}
\newcommand{\braket}[2]{\left<{#1}|{#2}\right>}
\newcommand{\ketbra}[2]{\left|{#1}\right>\left<{#2}\right|}
\newcommand{\commt}[2]{\left[{#1},{#2}\right]}
\newcommand{\tr}[1]{\mbox{Tr}{#1}}

\newcommand{\identity}{1\!\!1}

\begin{abstract}
We present a study on two-particle indistinguishability and particle-species identification by introducing a Fisher-information (FI) approach---in which two particles pass through a two-wave mixing operation and the number of particles is counted in one of the output modes. In our study, we first show that FI can reproduce the Hong-Ou-Mandel (HOM) effect with two bosons or two fermions. In particular, it is found that even though bosons and fermions exhibit different physical behavior (i.e., ``bunching" or ``anti-bunching") due to their indistinguishability, the aspects of HOM-like dip are quantitatively same. We then provide a simple method for estimating the degree of two-particle indistinguishability in a Mach-Zehnder interferometer-type setup. The presented method also enables us to identify whether the particles are bosons or fermions. Our study will provide useful primitives for various study of  boson and fermion characteristics.

\keywords{Fisher information \and Two-particle indistinguishability \and Particle-species identification}

\end{abstract}

\section{Introduction}

Fisher information (FI), originally introduced by R.A. Fisher~ \cite{Fisher}, is often regarded as a measure of indeterminacy and has been used to derive the lower bound for the mean-squared error of an unbiased estimator~\cite{CT2006}. The FI is exploited to determine a small deviation from a true value of a parameter with a fixed measurement~\cite{GLM11}. Given a specific measurement basis, we can readily evaluate the FI by processing the measurement outcomes. In quantum theory, on the other hand, the FI can be maximized by considering all possible measurement bases, and it presents quantum Fisher information (QFI) which is an intrinsic quantity for an input quantum state. 
The QFI was studied in quantum metrology, aiming at achieving better precision in a parameter estimation and enhancing its sensitivity \cite{GLM11,BC94,Oh19}. 
The QFI is widely used for various metrological tasks~\cite{Luo03,LWS10,FD12,KS13,Wang14,CJD17,Jarzyna17,PS2009,H2012,LL2013,HLS2015,S2016,KKW2018}, 
but this paper focuses on the (classical) FI to get a better insight in quantum phenomena and sensitivity of parameters. Except quantum metrology, previously there are only few quantum tasks studied by the classical FI~\cite{Hall00,YF17}.

The FI is related with the relative entropy, i.e., a measure of the statistical difference between two probability distributions $P(X)$ and $P(X_0)$~\cite{CT2006}.  For $\Delta X\equiv X-X_0 \ll 1$, the relative entropy, also called ``Kullback-Leibler divergence'', can be approximated as 
\begin{eqnarray}
D(P(X)||P(X_0))=\sum_y P(y|X)\ln\frac{P(y|X)}{P(y|X_0)}\approx \frac{(\Delta X )^2}{2}F(X_0),
\end{eqnarray}
where $P(y|X)$ and $P(y|X_0)$ denote the conditional probabilities of getting the measurement outcomes $y$ for the parameters $X$ and $X_0$. Here, the FI of $X_0$ is defined as
\begin{eqnarray} 
F(X_0)=\sum_y\frac{1}{P(y|X_0)}\bigg(\frac{\partial P(y|X_0)}{\partial X_0}\bigg)^2.
\label{eq:FI_def}
\end{eqnarray}
From the two neighborhood probability distributions, the relative entropy is thus approximated as FI of $X_0$, multiplied by $\frac{{\Delta X}^2}{2}$~\cite{BC94}.

\begin{figure}
\centering
\includegraphics[width=0.80\textwidth]{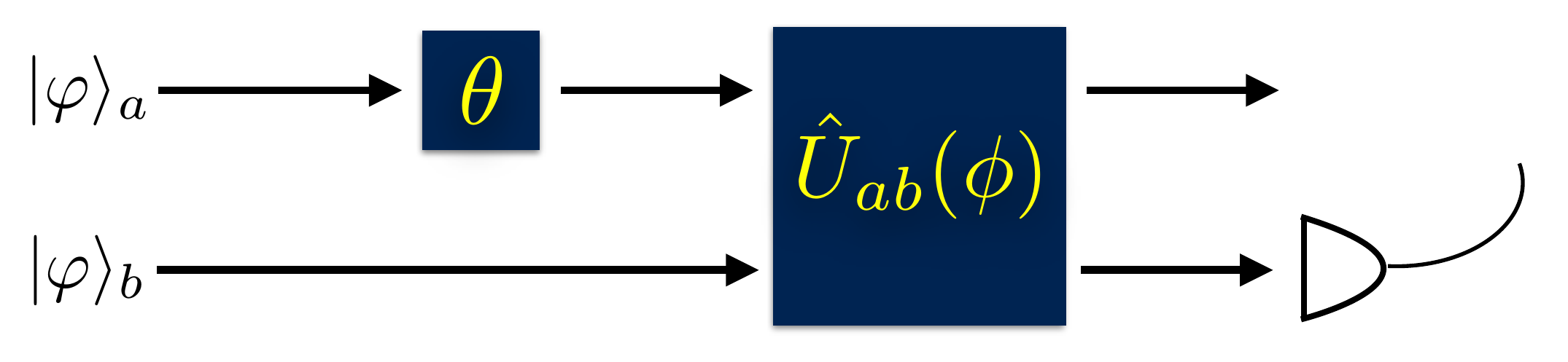}
\caption{\label{fig:fig1} A schematic picture of our Fisher-information (FI) approach. The two input particles in the states of $\ket{\varphi_1}_a$ and $\ket{\varphi_1}_a$ are fed into a two-wave mixing (TWM) operation, $\hat{U}_{ab}(\theta, \phi)$ with the parameters $\theta$ and $\phi$. After the interaction, we perform measurements repeatedly in one of the output modes to evaluate the FI.}
\end{figure}  

Using the above-described properties of FI and its relation to the relative entropy, we present a FI approach to yield useful primitives for understanding ``boson'' and ``fermion'' characteristics. Here, we utilize the indistinguishability of the identical particles. Due to the quantum interference of the identical particles, the particle indistinguishability was exploited to enhance quantum information protocols~\cite{FC18,id1,id2,id3,id4}. Our approach is based on the use of a two-wave mixing (TWM) operation, e.g., beam-splitter or Mach-Zehnder interferometer (MZI). Specifically, we feed the particles into a TWM operation, say $\hat{U}_\text{ab}$, and after passing through, the FI is evaluated by measuring only one of the output modes (see Fig.~\ref{fig:fig1}). Our study is two-fold: ($i$) Firstly, we show that FI exhibits the Hong-Ou-Mandel (HOM) effect with both bosons and fermions which exhibit the same FI. ($ii$) Secondly, we propose a simple method to estimate the degree of the two-particle indistinguishability, identifying their species, namely whether they are bosons or fermions. This is enabled by using the MZI-type operation. All these features are exhibited consistently for both bosons and fermions with different underlying physics, i.e., ``bunching'' and ``anti-bunching.'' 
Our study presents useful insight on the issues raised by other related works~\cite{FC18,BM18,KMKK18,KKK18}.

\section{Equivalence between global and local Fisher information}

We indicate that in our scenario (as depicted in Fig.~\ref{fig:fig1}) the evaluation of FI can be carried out by two different ways: i.e., global and local ways. Here,  the``global''  means that the FI is constructed by measuring both output modes. On the other hand, we can also evaluate the FI by counting the particle number in only one of the output modes. We define this ``local'' FI. In this section, we prove that the global and local FIs are quantitatively equivalent. Firstly, let us consider the output after a TWM operation on two particles. Without loss of the generality, the state of the output is given as
\begin{eqnarray}
a(\theta,\phi)|1,1\rangle_{ab}+b(\theta,\phi)|2,0\rangle_{ab}+c(\theta,\phi)|0,2\rangle_{ab},
\end{eqnarray}
where $|a(\theta,\phi)|^2+|b(\theta,\phi)|^2+|c(\theta,\phi)|^2=1$. 
The state of $|n,m\rangle_{ab}$ represents $n$ and $m$ particles in the two modes $a$ and $b$, respectively. 
$\theta$ and $\phi$ are the parameters of the TWM.
Subsequently, by considering the conditional probabilities
\begin{eqnarray}
P_{ab}(n_a,m_b|\theta,\phi) &=& \langle \hat{n}_a\otimes\hat{m}_b\rangle, \nonumber \\
P_{a}(n_a|\theta,\phi) &=& \langle \hat{n}_a\otimes\hat{\identity}_b\rangle, \nonumber \\
P_{b}(m_b|\theta,\phi) &=& \langle \hat{\identity}_a\otimes\hat{m}_b\rangle,
\end{eqnarray}
we can verify the following properties:
\begin{eqnarray}
\left\{
\begin{array}{l}
P_{ab}(1,1|\theta,\phi)=P_{a}(1|\theta,\phi)=P_{b}(1|\theta,\phi)=|a(\theta,\phi)|^2, \\ 
P_{ab}(2,0|\theta,\phi)=P_{a}(2|\theta,\phi)=P_{b}(0|\theta,\phi)=|b(\theta,\phi)|^2, \\
P_{ab}(0,2|\theta,\phi)=P_{a}(0|\theta,\phi)=P_{b}(2|\theta,\phi)=|c(\theta,\phi)|^2.
\end{array}
\right.
\end{eqnarray}
$P_{ab}(n_a,m_b|\theta,\phi)$ represents the probability of detecting $n$  and $m$ particles in the modes $a$ and $b$, respectively. 
$P_{a(b)}(n_{a(b)}|\theta,\phi)$ means the probability of detecting $n$ particles in the mode $a(b)$.
Then, we arrive at
\begin{eqnarray}
F_{ab}(\theta)=F_{a}(\theta)=F_{b}(\theta)~\text{and}~F_{ab}(\phi)=F_{a}(\phi)=F_{b}(\phi).
\label{eq:smm}
\end{eqnarray}
This feature can save the cost with less experimental efforts, because it allows us to explore the particle characteristics by the local FI taking into account the favorable parameter setting. Thus, hereafter we focus on the local FI.



\section{Reproducing Hong-Ou-Mandel effect with Fisher information}

Now we consider a scenario that a parameter is encoded in one input mode. A small difference between the two particles is quantifiable by the Hong-Ou-Mandel effect, which originally exhibits bunching effect of two bosonic particles~\cite{HOM87}. Initially, two identical single photons enter coherently into a 50:50 beam splitter, one on each side. 
The beam-splitting transformation is represented by 
\begin{eqnarray}
\hat{a}^{\dag} \rightarrow \frac{\hat{a}^{\dag}+\hat{b}^{\dag}}{\sqrt{2}}~\text{and}~\hat{b}^{\dag}\rightarrow \frac{\hat{b}^{\dag}-\hat{a}^{\dag}}{\sqrt{2}}, 
\end{eqnarray}
where $\hat{a}$ and $\hat{a}^\dagger$ ($\hat{b}$ and $\hat{b}^\dagger$) are bosonic annihilation and creation operators in mode $a$ (mode $b$), respectively. Using this beam-splitting transformation, we can have a two-mode output state 
\begin{eqnarray}
\hat{B}_{ab}\ket{1, 1}_{ab}=\frac{\ket{0,2}_{ab}-\ket{2,0}_{ab}}{\sqrt{2}}.
\end{eqnarray}
Given a time delay between the two input single photons, we have a half probability of counting each photon in both output modes. With no time delay, the each photon counting event drops into zero by bunching effect. Using the probability of each photon counting event, we can observe HOM-like dip as a function of the time delay between the two input single photons. The indistinguishability of the two single photons determines the depth of the dip, which is bounded between zero and the maximum number of counts.

 \begin{figure}
\centering
\includegraphics[width=0.95\textwidth]{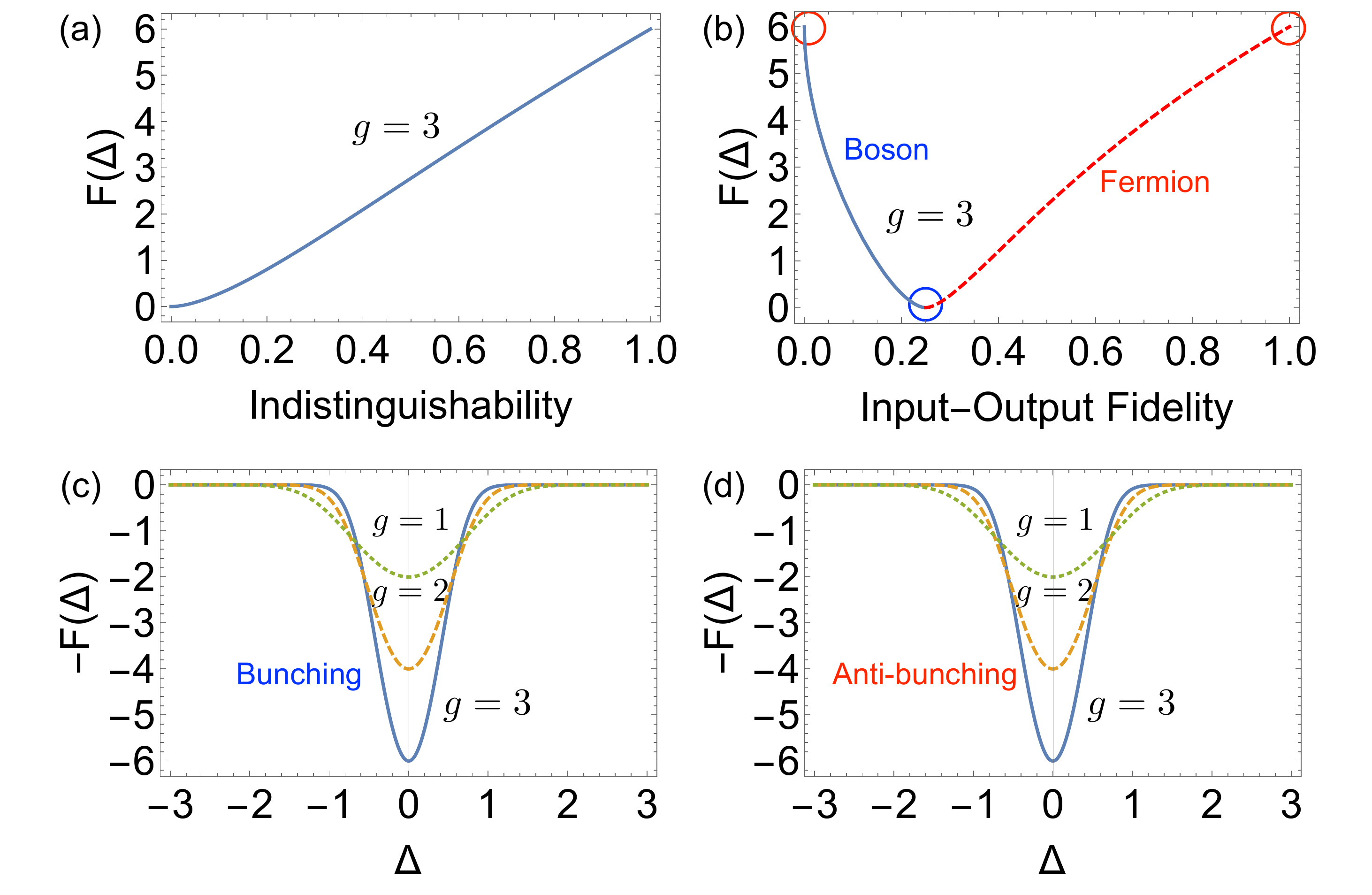}
\caption{\label{fig:fig2} We present the FI with respect to the control parameter $\Delta$. (a) The FI as a function of indistinguishability $\abs{\braket{\varphi_1}{\varphi_2}}^2$ of the input bosonic and fermionic states. (b) Relation between the FI and the input-output fidelity $\abs{\bra{\varphi_1,\varphi_2}\hat{B}_{ab}\ket{\varphi_1,\varphi_2}}^2$. The blue (red) circle indicates that the particles are completely distinguishable (indistinguishable). With these properties, the Hong-Ou-Mandel-like dips are reproduced, consistently for both (c) bosons and (d) fermions. Here, the factor $g$ determines the depth of the dip. Note that these aspects arise with different physical phenomena, i.e., ``bunching'' for bosons and ``anti-bunching'' for fermions.}
\end{figure} 

We illustrate that the original HOM effect can be observed with bosons and fermions by means of the FI. Firstly, let us consider the bosonic case. We replace the time delay by the polarization degree-of-freedom which exhibits the indistinguishability of the photons. We then introduce a control parameter, say $\Delta$, in order to control the HOM-like dip. Note that such an adoption of the polarization degree-of-freedom with $\Delta$ is physical and does not restrict the generality of our analysis. In this setting, two polarized photon states $\ket{\varphi_1}_a=\sqrt{e^{-g\Delta^2}}\ket{H}_a+\sqrt{1-e^{-g\Delta^2}}\ket{V}_a$ and $\ket{\varphi_2}_b=\ket{H}_b$ are fed into a 50:50 beam splitter as
\begin{eqnarray}
\hat{B}_{ab}\left(\sqrt{e^{-g\Delta^2}}\ket{H}_a+\sqrt{1-e^{-g\Delta^2}}\ket{V}_a\right)\ket{H}_b,
\end{eqnarray}
where a factor $g$ is related to the length of the photon wave packet in the original experiment. By counting the number of photons in the output mode $b$, the photon-number probability distributions are characterized as
\begin{eqnarray}
P_b(0|\Delta) &=& P_b(2|\Delta)=\frac{1}{4}(1+e^{-g\Delta^2}), \nonumber \\
P_b(1|\Delta) &=& \frac{1}{2}(1-e^{-g\Delta^2}).
\end{eqnarray}
Consequently, from Eq.~(\ref{eq:FI_def}), the FI of $\Delta$ can be evaluated as
\begin{eqnarray}
F(\Delta)=\frac{4g^2\Delta^2}{e^{2g\Delta^2}-1},
\label{eq:FI_delta_B}
\end{eqnarray}
where $0 \le F(\Delta) \le 2g$. 
The minimum and the maximum values are obtained with different and same polarized states, respectively. Specifically, the value of $F(\Delta)$ exactly represents `indistinguishability' of the two input photons---here, the indistinguishability is defined as the fidelity between the two input particle states $\abs{\braket{\varphi_1}{\varphi_2}}^2$. This feature is equivalent to that of the original HOM effect. In Fig.~\ref{fig:fig2}(a), we depict the graph of $F(\Delta)$ with respect to the indistinguishability of the two input photons for $g=3$. Additionally, we also investigate the relation between $F(\Delta)$ and an input-output fidelity $\abs{\bra{\varphi_1,\varphi_2}\hat{B}_{ab}\ket{\varphi_1,\varphi_2}}^2$. It is found that $F(\Delta)$ is {\em inversely proportional} to the input-output fidelity (blue solid curve in Fig.~\ref{fig:fig2}(b)). Obviously, this is attributed to the `bunching' effect of the bosonic system. Based on the above results, we can produce HOM-like dip, by taking minus sign on $F(\Delta)$ (see Fig.~\ref{fig:fig2}(c)). Here, the factor $g$ determines the depth of the dip.

The same behaviors of the HOM effect can occur even in fermionic systems. In this case, two electron-spin states $\ket{\varphi_1}_a=\sqrt{e^{-g\Delta^2}}\ket{\downarrow}_a+\sqrt{1-e^{-g\Delta^2}}\ket{\uparrow}_a$ and $\ket{\varphi_2}_b=\ket{\downarrow}_b$ are fed into a 50:50 electronic beam-splitter as 
\begin{eqnarray}
\hat{B}^e_{ab}\left(\sqrt{e^{-g\Delta^2}}\ket{\downarrow}_a+\sqrt{1-e^{-g\Delta^2}}\ket{\uparrow}_a\right)\ket{\downarrow}_b,
\end{eqnarray} 
where the electronic beam splitting transformation is defined by \cite{Loudon1998}
\begin{eqnarray}
\hat{f}^{\dag}_a \rightarrow \frac{\hat{f}^{\dag}_a+\hat{f}^{\dag}_b}{\sqrt{2}}~~\text{and}~~\hat{f}^{\dag}_b \rightarrow \frac{\hat{f}^{\dag}_b-\hat{f}^{\dag}_a}{\sqrt{2}}.
\end{eqnarray}
$\hat{f}_a$ and $\hat{f}_a^\dagger$ ($\hat{f}_b$ and $\hat{f}_b^\dagger$) are fermionic annihilation and creation operators in mode $a$ (mode $b$), respectively. Here, the control parameter $\Delta$ and the factor $g$ are for the single-electron state, and they play the same role as the parameters of the bosonic case. By counting the number of electrons in the output mode $b$, we can characterize the probability distributions as
\begin{eqnarray}
P_b(0|\Delta)&=&P_b(2|\Delta)=\frac{1}{4}(1-e^{-g\Delta^2}), \nonumber \\
P_b(1|\Delta)&=&\frac{1}{2}(1+e^{-g\Delta^2}),
\end{eqnarray}
and then the corresponding FI of $\Delta$ exhibits the same formula as the Eq.~(\ref{eq:FI_delta_B}) of the bosonic case. Thus, $F(\Delta)$ evaluated by the fermions increases monotonically with respect to the indistinguishability of two electron states,  as shown in Fig.~\ref{fig:fig2}(a). However, it is {\em proportional} to the input-output fidelity (red dashed curve in Fig.~\ref{fig:fig2}(b)), contrary to the case of bosons. Actually, this is attributed to the `anti-bunching' effect, which is a property of fermionic system, namely from the Pauli exclusion principle. Thus, the HOM-like dip also appears near the completely same electron-spin states (see Fig. \ref{fig:fig2}(d)), even though the underlying physics is complete different from that of bosons. 

\section{Measure of indistinguishability and particle species}

We consider a parameter of a two-wave mixing operation: specifically, a phase-shifting parameter in a MZI-type operation, which is defined as
\begin{eqnarray} 
\hat{U}_{MZI}=\hat{B}_{ab}e^{i\phi\hat{a}^{\dag}\hat{a}}\hat{B}_{ab},
\end{eqnarray}
where $e^{i\phi\hat{a}^{\dag}\hat{a}}$ is a single-mode operation for shifting a phase $\phi \in (0, \pi]$\footnote{Here we do not consider the case of $\phi=0$.}. This MZI operation is defined consistently for both bosons and fermions~\cite{Loudon1998,eMZI}.

\begin{figure}
\centering
\includegraphics[width=0.95\textwidth]{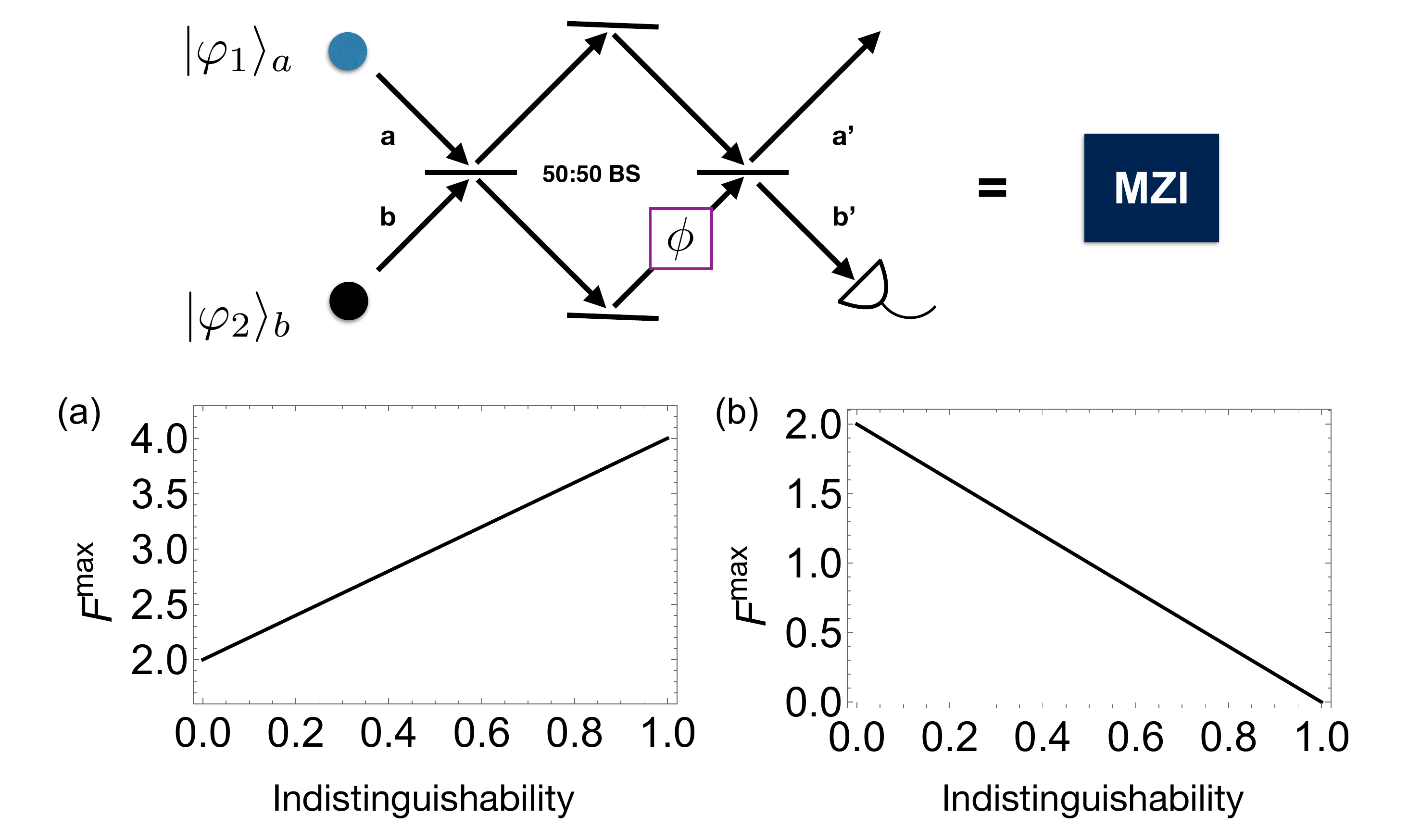}
\caption{\label{fig:fig3} A simple framework to study the two-particle indistinguishability ($\abs{\braket{\varphi_1}{\varphi_2}}^2 = \abs{\beta}^2$ in our case) by using the Mach-Zehnder interferometer (MZI) operation. We calculate the maximum value of FI, say $F^\text{max}$, with respect to the phase $\phi$: i.e., $F^\text{max} = \max_{\phi}F(\phi)$ for both bosons and fermions. As the results, we present the relation between $F^\text{max}$ and the indistinguishability of (a) bosons and (b) fermions.}
\end{figure} 

Firstly, let us consider the bosonic case, where we use two photons---one is assumed to be in the state of $\ket{\varphi_1}_a=\sqrt{1-\abs{\beta}^2}\ket{V}_a + \beta\ket{H}_a$ and the other is in $\ket{\varphi_2}_b=\ket{H}_b$. Then, we can obtain the probability distributions $P(n_{H_{a'}},m_{V_{a'}}; k_{H_{b'}}, l_{V_{b'}})$ by counting the number of photons and identifying their polarization, where $n_{H_{a'}}$ and $m_{V_{a'}}$ (or $k_{H_{b'}}$ and $l_{V_{b'}}$) are the number of $H$-polarized and $V$-polarized photons in the output $a'$-mode (or $b'$-mode), respectively.  
However, owing to Eq.~(\ref{eq:smm}), it is enough to characterize the following conditional probabilities:
\begin{eqnarray}
P_{b'}(1_H|\phi)&=&\frac{1-\abs{\beta}^2}{16}\abs{1-e^{i\phi}}^4+\frac{\abs{\beta}^2}{4}\abs{1+e^{i 2\phi}}^2, \nonumber \\
P_{b'}(1_V|\phi)&=&\frac{1-\abs{\beta}^2}{16}\abs{1+e^{i\phi}}^4, \nonumber \\
P_{b'}(2_H|\phi)&=&\frac{\abs{\beta}^2}{8}\abs{1-e^{i 2\phi}}^2, \nonumber \\
P_{b'}(1_H,1_V|\phi)&=&\frac{1-\abs{\beta}^2}{16}\abs{1-e^{i 2\phi}}^2,
\end{eqnarray}
where $P_{b'}(0|\phi) = P_{b'}(2_H|\phi)+P_{b'}(1_H,1_V|\phi)$. With these probability distributions, we can directly obtain the FI of the phase $\phi$ as the function of the indistinguishability, i.e., $\abs{\braket{\varphi_2}{\varphi_1}}^2=\abs{\beta}^2$, such that
\begin{eqnarray}
F(\phi) &=& \left( 1-\abs{\beta}^2 \right) + \left( 1+3\abs{\beta}^2 \right) \cos^2{\phi} \nonumber \\
           && +\frac{\left[ 2\left(1-\abs{\beta}^2\right)\sin{\phi} - \left(1+3\abs{\beta}^2 \right)\sin{2\phi} \right]^2}{4\left( 1-\abs{\beta}^2 \right)\left( 1-\cos{\phi} \right)^2 + 8\abs{\beta}^2\left( 1+\cos{2\phi} \right)}.
\end{eqnarray}
For a given $\abs{\beta}^2$, let us define the maximum value of $F(\phi)$ over $\phi$ as $F^\text{max} \equiv \max_{\phi \in (0, \pi]}{F(\phi)}$. Then, it is found that
\begin{eqnarray}
F^\text{max} = 2 \left( 1 + \abs{\beta}^2 \right)~\text{for}~\phi = \pi,
\end{eqnarray}
which indicates that $F^\text{max}$ monotonically increases from $2$ up to $4$ with respect to the indistinguishability of the two bosonic particles (see Fig.~\ref{fig:fig3}(a)). This feature is due to the bunching property of the bosons.

Secondly, in the fermionic case, we feed two electrons into the electronic MZI operation~\cite{eMZI}. Here, the two electrons are assumed to be in the states of $\ket{\varphi_1}_a=\sqrt{1-\abs{\beta}^2}\ket{\uparrow}_a + \beta\ket{\downarrow}_a$ and $\ket{\varphi_2}_b = \ket{\downarrow}_b$, respectively.  The probability distributions $P(n_{\uparrow_{a'}},m_{\downarrow_{a'}}; k_{\uparrow_{b'}}, l_{\downarrow_{b'}})$ are also defined by counting the number of electrons and identifying the spin. Here, in order to evaluate FI, we can also characterize the useful conditional probabilities:
\begin{eqnarray}
P_{b'}(1_{\uparrow}|\phi)&=&\frac{1-\abs{\beta}^2}{16}\abs{1+e^{i \phi}}^4, \nonumber \\
P_{b'}(1_{\downarrow}|\phi)&=&\abs{\beta}^2+\frac{1-\abs{\beta}^2}{16}\abs{1-e^{i\phi}}^4, \nonumber \\
P_{b'}(1_{\uparrow},1_{\downarrow}|\phi)&=&\frac{1-\abs{\beta}^2}{16}\abs{1-e^{i2\phi}}^2=P_{b'}(0|\phi).
\end{eqnarray}
Using these probability distributions, we can obtain the FI of the phase $\phi$ as
\begin{eqnarray}
F(\phi)=\left( 1- \abs{\beta}^2 \right)\left(1+\cos^2{\phi}\right)+\frac{\left( 1- \abs{\beta}^2 \right)\left(1-\cos{\phi}\right)^2\sin^2{\phi}}{\left(1-\cos{\phi}\right)^2+ \frac{4\abs{\beta}^2}{1- \abs{\beta}^2}}.
\end{eqnarray}
With the above-obtained $F(\phi)$, we can define $F^\text{max} = \max_{\phi \in (0, \pi]}{F(\phi)}$, and it is found that
\begin{eqnarray}
F^\text{max} = 2 \left( 1 - \abs{\beta}^2 \right)~\text{for}~\phi = \pi.
\end{eqnarray}
On the contrary to the bosonic case, it is observed that the value of $F^\text{max}$ monotonically decreases from $2$ to $0$ with respect to the indistinguishability of the two fermionic particles (see Fig.~\ref{fig:fig3}(b)). This is caused by the anti-bunching property of the fermions. 

\begin{figure}
\centering
\includegraphics[width=0.85\textwidth]{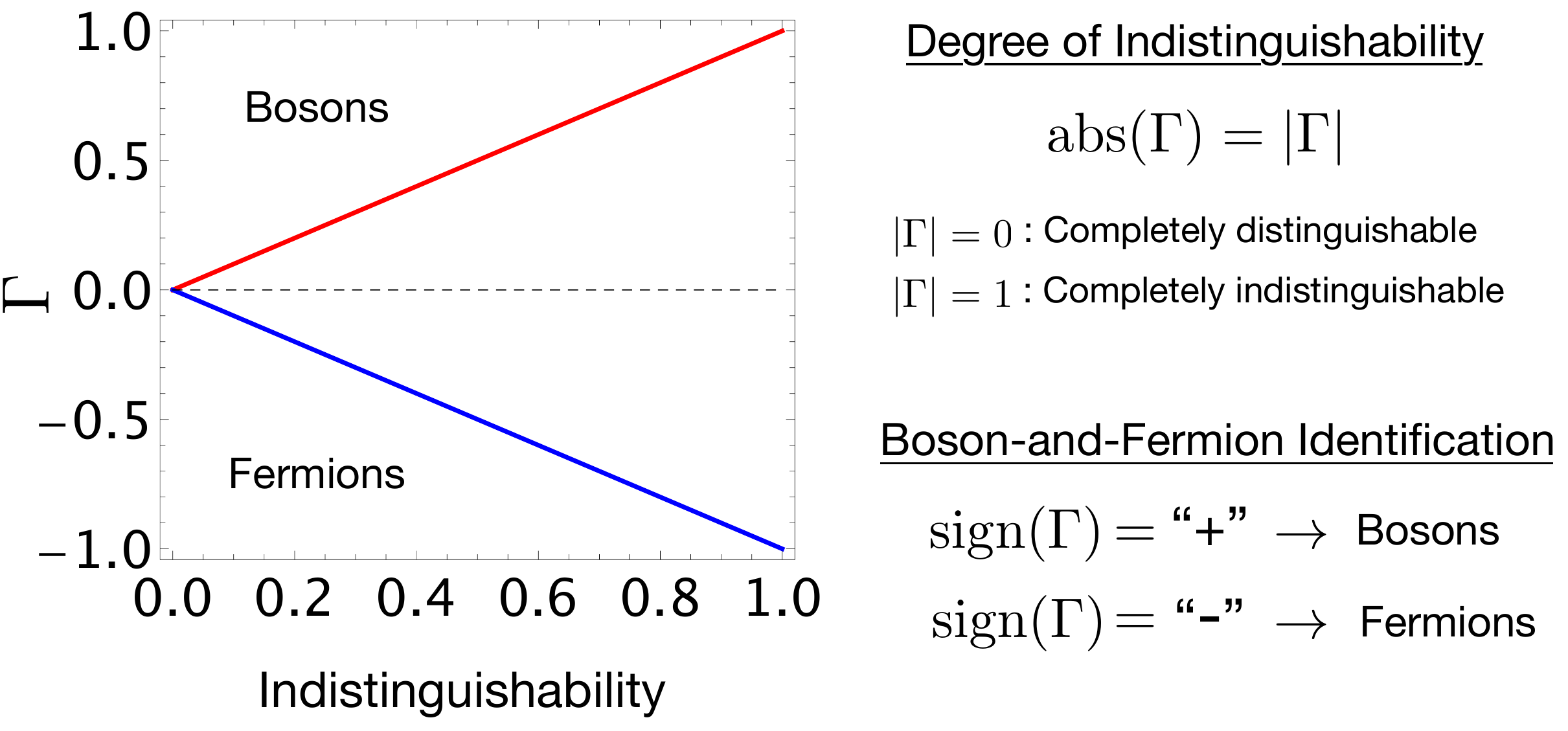}
\caption{\label{fig:fig4} We define a useful quantity $\Gamma$, as Eq.~(\ref{eq:Gamma}). This quantity $\Gamma$ allows us to estimate the degree of indistinguishability and the particle species by evaluating its absolute value and sign.}
\end{figure} 

On the basis of the above observations, we define a quantity\footnote{We note that such a definition by using the maximum FI occurs, frequently, in some physical contexts (see Refs. \cite{Fisher,FC96}).}
\begin{eqnarray}
\Gamma \equiv \frac{1}{2} \max_{\phi}{F(\phi)} - 1.
\label{eq:Gamma}
\end{eqnarray}
which is bounded as $-1 \le \Gamma \le 1$. This quantity $\Gamma$ is indeed useful. First, we can estimate the two-particle indistinguishability; specifically, for both bosons and fermions, we draw a new definition with the absolute value of $\Gamma$:
\begin{eqnarray}
\text{Degree of indistinguishability} \equiv \text{abs}(\Gamma). 
\end{eqnarray}
Here, the particles are completely indistinguishable (or distinguishable) when $\text{abs}(\Gamma)=1$ (or $\text{abs}(\Gamma)=0$). Furthermore, in our approach, the question of whether the (unknown) input particles are bosons or fermions can also be answered by evaluating $\text{sign}(\Gamma) \in \{+, -\}$; namely, we can identify the particle species, such that
\begin{eqnarray}
\left\{
\begin{array}{ll}
\text{Boson}    &~\text{when}~0 < \Gamma \le 1, \\
\text{Fermion} &~\text{when}~-1 \le \Gamma < 0.
\end{array}
\right.
\end{eqnarray}
This feature originates from the different properties (bunching and anti-bunching) of the particle species. Currently, developing different methods of assessing the indistinguishability is a growing interest, as potentially it could be useful as a resource of quantum tasks \cite{FC18,BM18,KMKK18,KKK18,KPK2018}. 

Before closing, we briefly note that one may raise a question of how to distinguish a pure input state $\sqrt{1-\abs{\beta}^2}\ket{V}_a+\beta\ket{H}_a$ from a mixed state $\left(1-\abs{\beta}^2\right)\ket{V}_a\bra{V}+\abs{\beta}^2\ket{H}_a\bra{H}$. Actually, such a task is possible by adjusting the state of the other input mode $b$; specifically, by changing the parameter $\theta$ in the state of $\cos{\theta}\ket{H}_b + \sin{\theta}\ket{V}_b$, we can identify the (degree of) interference term in the state, and hence, can evaluate which of these, i.e., purity or mixedness, is relevant. We defer the details to Appendix~A.



\section{Summary and Discussion}

We have presented a study on boson-and-fermion characteristics by using FI. Our FI approach was developed under a two-wave mixing (TWM) operation (as depicted in Fig.~\ref{fig:fig1}). Using the symmetric behavior of TWM operation, we investigated the particle characteristics by measuring only one of the output modes and taking into account the parameter(s) that we are interested in. We firstly showed that the FI can reproduce the Hong-Ou-Mandel-like effect, consistently for both bosons and fermions. Under the degree of indistinguishability of the input particles, however, bosons and fermions are related to different physical phenomena, namely of the bunching and anti-bunching, respectively. We then proposed a simple method to estimate the degree of two-particle indistinguishability and to identify their particle species in the MZI-type setup. It is expected that our study will provide different viewpoints in terms of the FI and be extended to many-particle scenarios. For many-particle systems, we may take a symmetric transformation that can present a relation between the local and global FIs. It will be a good candidate to consider multi-port beam splitters~\cite{ZZH97}, including the idea of an integrated tritter~\cite{sp18}.

\begin{acknowledgements}
S.-Y. L. and J.B. would like to thank Changhyoup Lee, Jaeyoon Cho, Tomasz Paterek, Junghee Ryu, and Marcin Wie\'sniak for useful discussions and comments. We acknowledge the financial support of the Basic Science Research Program through the National Research Foundation of Korea (NRF) (No. 2018R1D1A1B07048633).
\end{acknowledgements}

\appendix{\bf Appendix A: Distinguishing an input pure state from an input mixed state by Fisher information} \\

\begin{figure}
\centering
\includegraphics[width=0.47\textwidth]{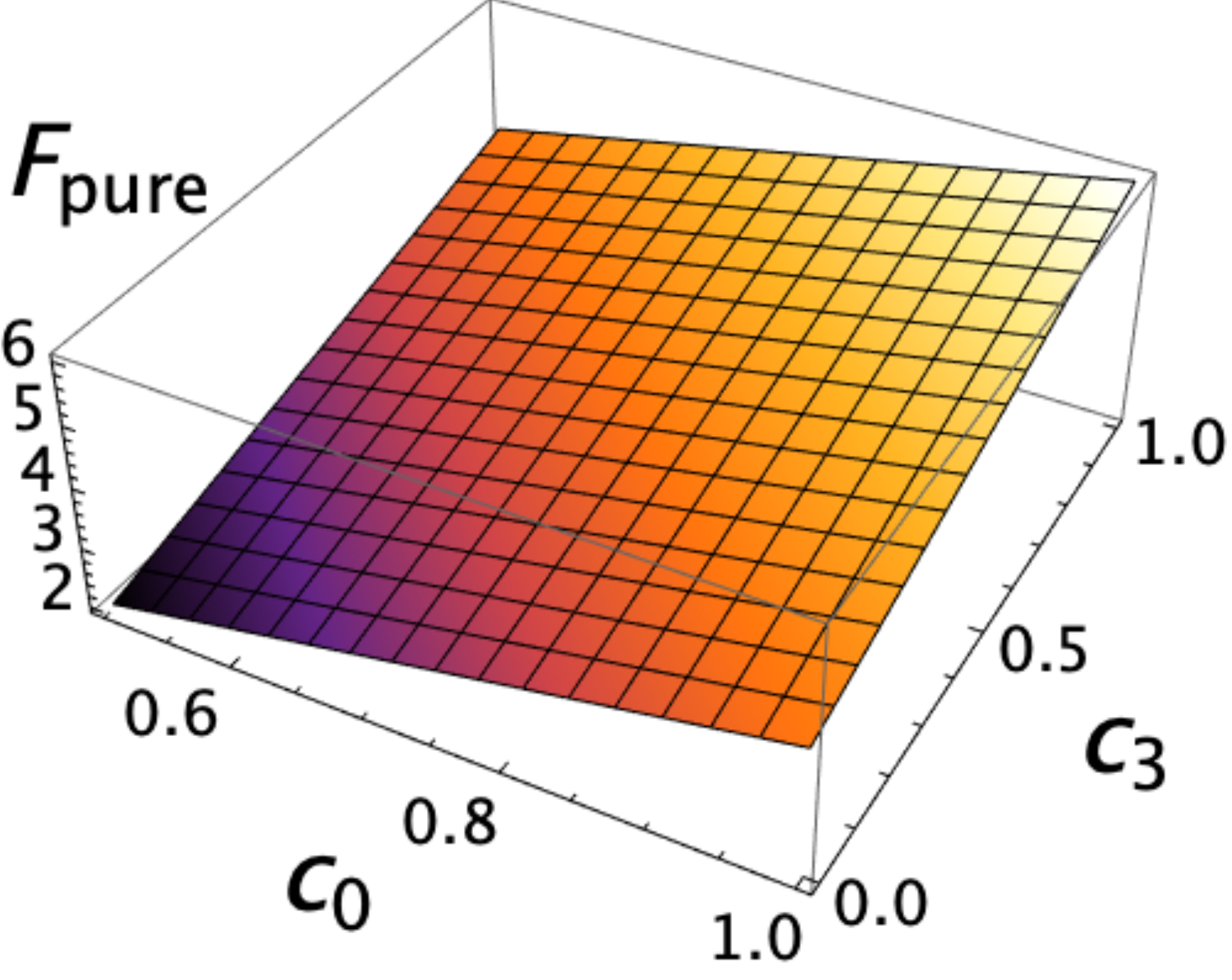}
\includegraphics[width=0.47\textwidth]{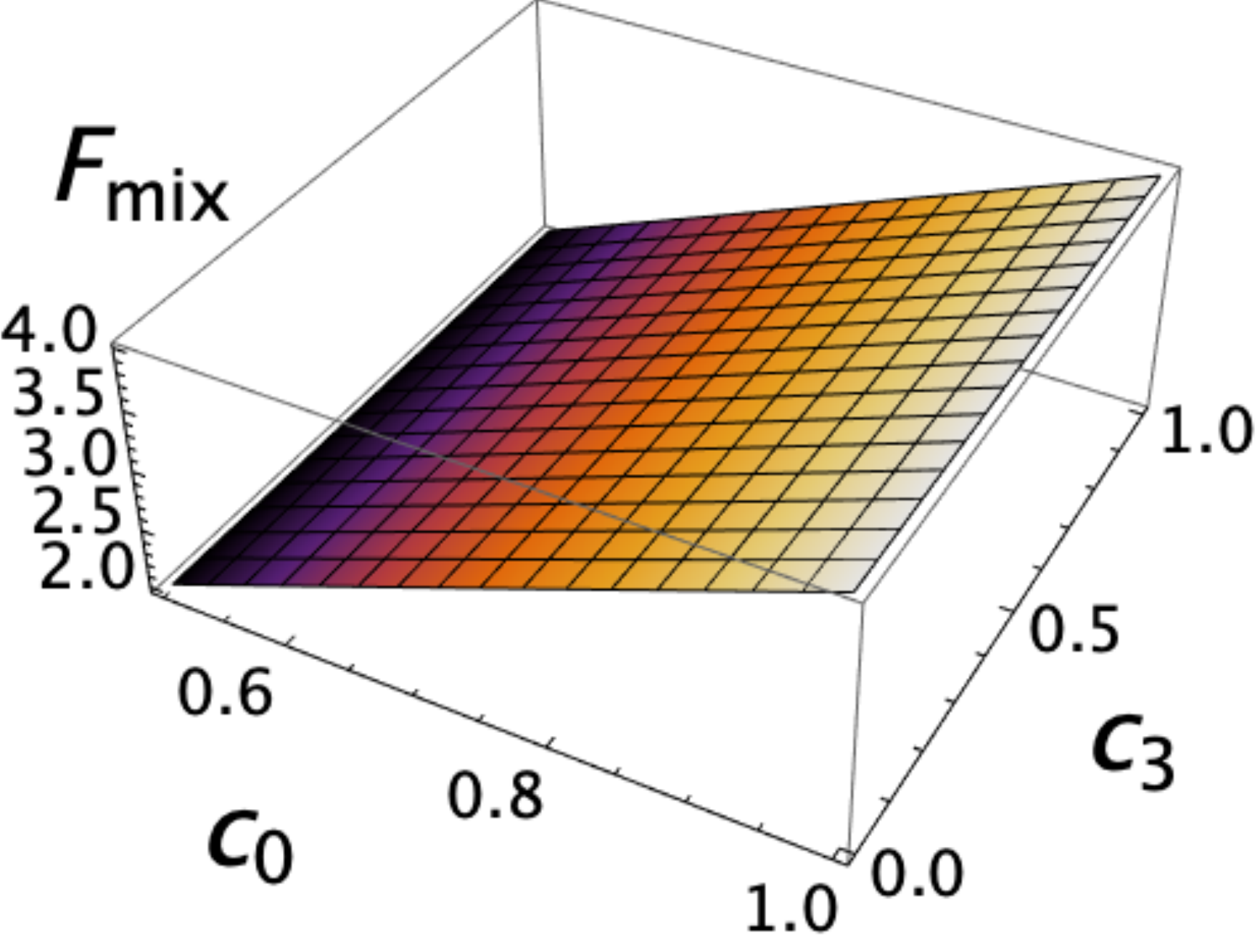}
\caption{\label{fig:fig5_appendixA}  The graphs of (left) $F_{\text{pure}}(\phi)$ and (right) $F_{\text{mix}}(\phi)$ evaluated for the pure and mixed bosonic input states in $a$-mode. Here, the phase $\phi$ is set to be $0.1$.}
\end{figure} 

For simplicity, here we count the number of particles without discriminating their polarizations or spins. For bosons, we start with the single-photon of mode $b$ in the superposed state of $\cos{\theta}\ket{V}_b+\sin{\theta}\ket{H}_b$. When the single-photon is in a pure state $\alpha\ket{V}_a+\beta\ket{H}_a$, we obtain the probability distributions by counting the number of photons only in the output mode $b'$,  such that:
\begin{eqnarray}
P_{b'}(0|\phi)&=&P_{b'}(2|\phi)=\frac{1-\cos{2\theta}}{4}\left(c_0+\frac{c_3}{2}\right), \nonumber \\
P_{b'}(1|\phi)&=&\frac{1}{2}\left( 1 + c_1 \cos{2\phi}+ c_2\cos^2{\phi}-c_3\right). 
\end{eqnarray}
With the above probability distributions, we can have the FI of the phase parameter $\phi$ as
\begin{eqnarray}
F_{\text{pure}}(\phi)=2\left( c_0 + \frac{c_3}{2} \right)^2 \sin^2{2\phi}\left[ \frac{1}{\left(c_0+\frac{c_3}{2}\right)\left(1-\cos{2\phi}\right)}+\frac{1}{2-2\left(c_0+\frac{c_3}{2}\right)\sin^2{\phi}}\right],
\label{eq:Fb_pure}
\end{eqnarray}
where $c_0=\frac{1}{2}(1+c_1)$, $c_1=\abs{\alpha}^2\cos^2{\theta}+\abs{\beta}^2\sin^2{\theta}$, $c_2=\abs{\alpha}^2\sin^2{\theta}+\abs{\beta}^2\cos^2{\theta}$, and 
$c_3=(\alpha\beta^{\ast}+\beta\alpha^{\ast})\sin{\theta}\cos{\theta}$.  Here, note that the factor $c_3$ is related to the interference, and hence the purity, of the input state of the mode $a$. On the other hand, for a mixed state $\abs{\alpha}^2\ket{V}_a\bra{V}+\abs{\beta}^2\ket{H}_a\bra{H}$, the FI is given by
\begin{eqnarray}
F_{\text{mix}}(\phi)=2c_0^2 \sin^2{2\phi}\left[\frac{1}{c_0 \left(1-\cos{2\phi}\right)} +\frac{1}{2-2c_0 \sin^2{\phi}}\right].
\label{eq:Fb_mix}
\end{eqnarray}
 Consequently, we can figure out the difference between Eq.~(\ref{eq:Fb_pure}) and Eq.~(\ref{eq:Fb_mix}) in terms of the existence of the interference with $c_3$ (as long as $\phi \neq \frac{\pi}{2}$ and $\phi \neq \pi$) As an example, in Fig.~\ref{fig:fig5_appendixA} we depict the 3D-graphs of (left) $F_{\text{pure}}(\phi)$ and (right) $F_{\text{mix}}(\phi)$ for $\phi=0.1$.

\begin{figure}
\centering
\includegraphics[width=0.47\textwidth]{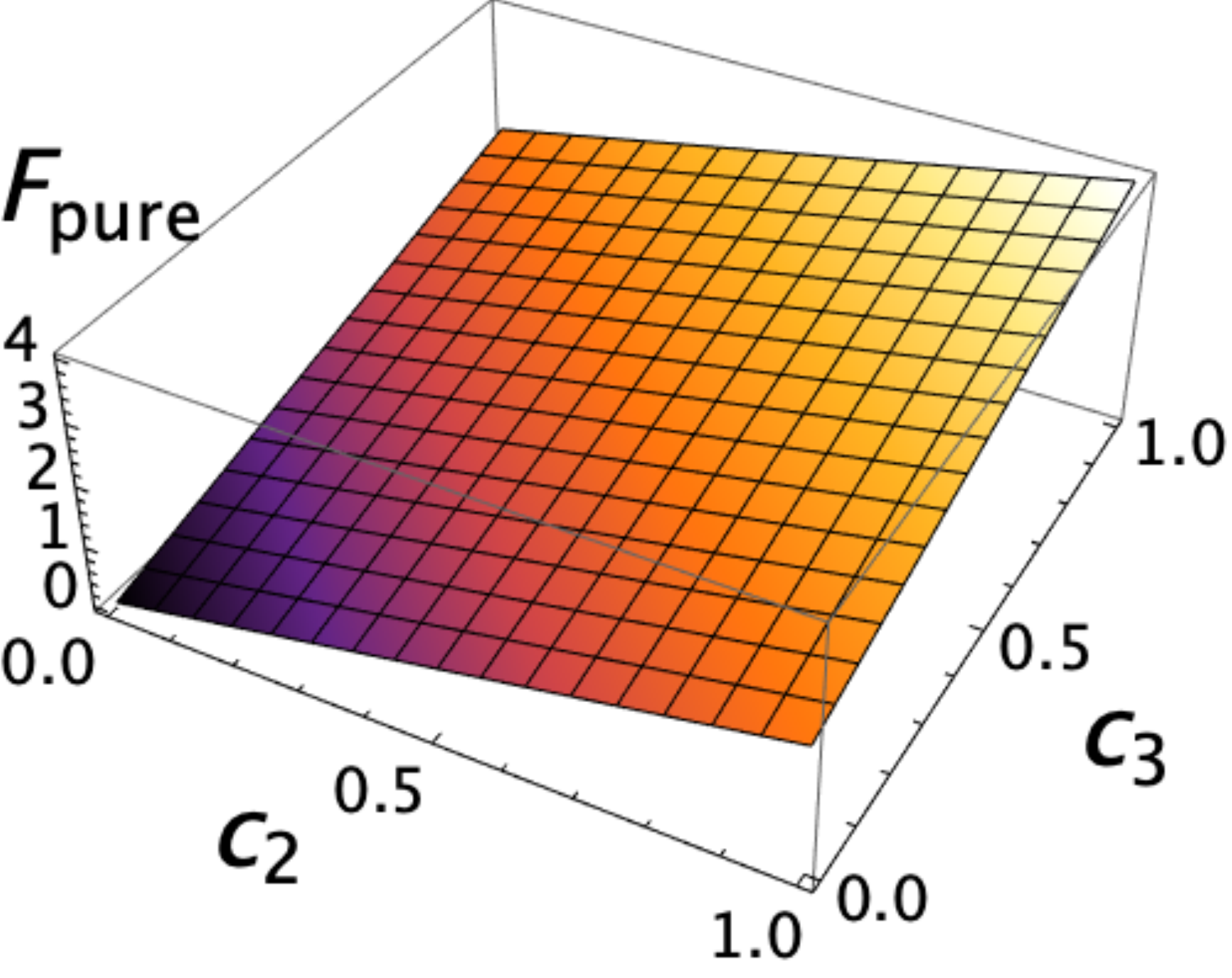}
\includegraphics[width=0.47\textwidth]{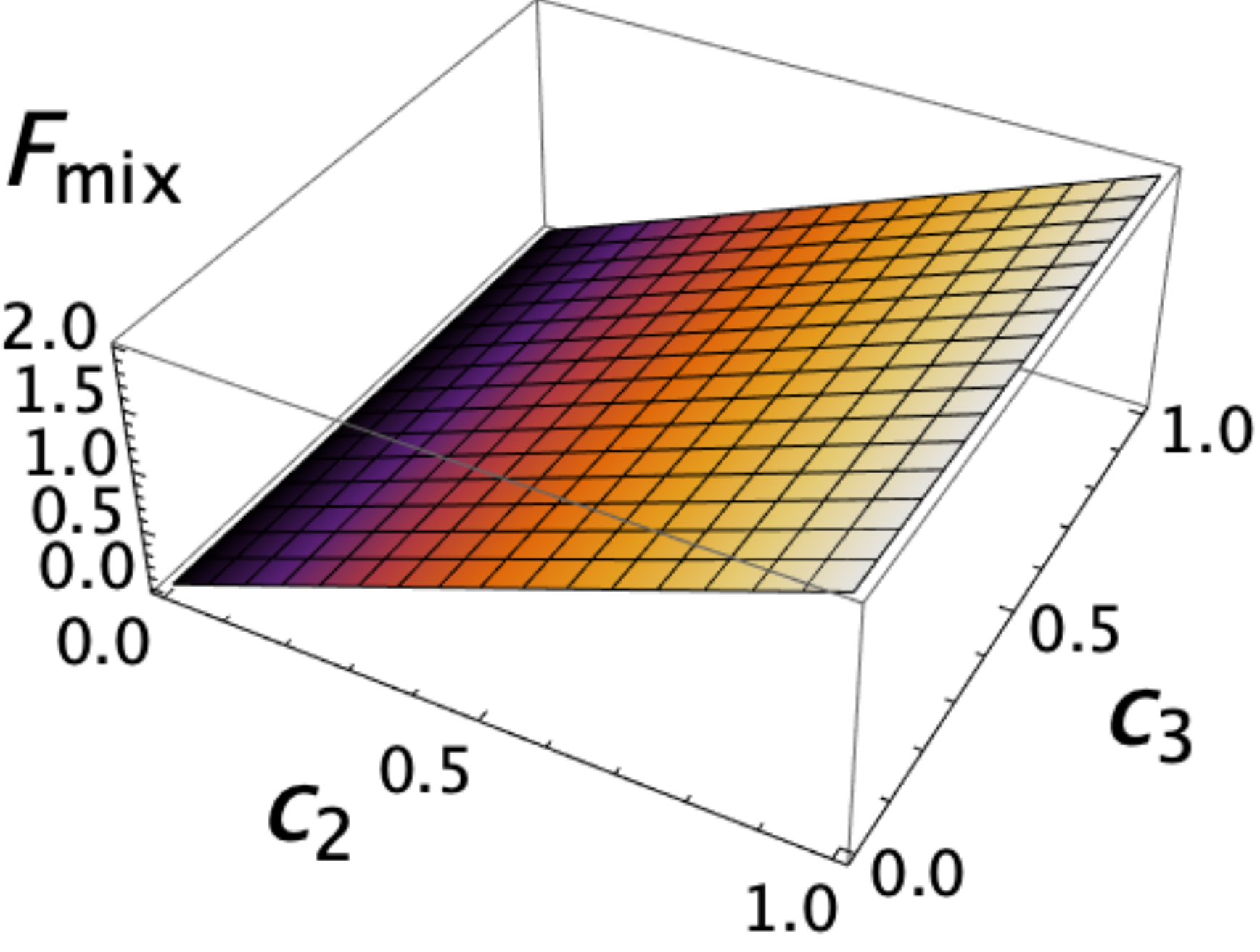}
\caption{\label{fig:fig6_appendixA}  The graphs of (left) $F_{\text{pure}}(\phi)$ and (right) $F_{\text{mix}}(\phi)$ evaluated for the pure and mixed fermionic input states in $a$-mode. Here, we also set $\phi = 0.1$.}
\end{figure} 

For fermions, we start with the single-electron of mode $b$ in the superposed state of $\cos{\theta}\ket{\uparrow}_b+\sin{\theta}\ket{\downarrow}_b$. When the single-electron of mode $a$ is in a pure state of $\alpha\ket{\uparrow}_a+\beta\ket{\downarrow}_a$, we obtain the probability distributions by counting the number of electrons only in the output mode $b'$,  such that:
\begin{eqnarray}
P_{b'}(0|\phi)&=&P_{b'}(2|\phi)=\frac{1-\cos{2\theta}}{8}\left(c_2+c_3\right), \nonumber \\
P_{b'}(1|\phi)&=&c_0+\frac{c_2 \cos^2{\phi}}{2} - \frac{c_3\sin^2{\phi}}{2}.
\end{eqnarray} 
Then, from the above probabilities, the FI of $\phi$ is given as
\begin{eqnarray}
F_{\text{pure}}(\phi)=\left(c_2+c_3\right)^2\sin^2{2\phi}\left[\frac{1}{\left(c_2+c_3\right)\left(1-\cos{2\phi}\right)}+\frac{1}{4-2\left(c_2+c_3\right)\sin^2{\phi}}\right].
\label{eq:Fp_pure}
\end{eqnarray}
On the other hand, for a mixed state $\abs{\alpha}^2\ket{\uparrow}_a\bra{\uparrow}+\abs{\beta}^2\ket{\downarrow}_a\bra{\downarrow}$ of the single-electron, the FI is given as
\begin{eqnarray}
F_{\text{mix}}(\phi)=c_2^2\sin^2{2\phi}\left[\frac{1}{c_2\left(1-\cos{2\phi}\right)}+\frac{1}{4-2c_2\sin^2{\phi}}\right].
\label{eq:Fp_mix}
\end{eqnarray}
As such, we can also figure out the difference of FIs in Eq.~(\ref{eq:Fp_pure}) and Eq.~(\ref{eq:Fp_mix}) with $c_3$ (as long as $\phi \neq \frac{\pi}{2}$ and $\phi \neq \pi$) In Fig.~\ref{fig:fig6_appendixA}, we also present the graphs of (left) $F_{\text{pure}}(\phi)$ and (right) $F_{\text{mix}}(\phi)$ for $\phi=0.1$, which exhibits the same behavior as the case of boson.\footnote{Note that $c_2~(\text{and}~c_1) \in [0,1]$, whereas $c_0 \in \left[\frac{1}{2}, 1\right]$.}

\end{document}